\documentclass[aps,prb,twocolumn,showpacs,amsmath,amssymb,longbibliography,10pt]{revtex4-2}

\usepackage{lineno}
\usepackage[utf8]{inputenc}
\usepackage{tgtermes}
\usepackage{bm}
\usepackage{color}
\usepackage{makeidx}
\usepackage{amsfonts}
\usepackage{amssymb}
\usepackage{amsthm}
\usepackage{amsmath}
\usepackage{graphicx}
\usepackage{epstopdf}
\usepackage{epsfig}
\usepackage{framed}
\usepackage{booktabs}
\usepackage[normalem]{ulem}
\usepackage{bbold}
\usepackage{multirow}
\usepackage[colorlinks=true,citecolor=blue,urlcolor=blue,linkcolor=blue]{hyperref}
\usepackage[dvipsnames,svgnames,x11names]{xcolor}
\usepackage{subfigure}
\usepackage{framed}
\usepackage{booktabs}
\usepackage{nicematrix}
\usepackage{tikz}
\usetikzlibrary{fit}
\tikzset{highlight/.style={rectangle,
        fill=red!15,
        rounded corners = 0.5 mm,
        inner sep=1pt,
        fit=#1}
        }
\begin{document}
	
\title{Efficient MPO Construction for Long-Range Hamiltonians with Periodic Boundary Conditions: Application to Many-Body Dynamics}
	
\author{Xin Wang}
\affiliation{Department of Physics, Wuhan University of Technology, Wuhan 430070, China}
\author{Bo Xiong}
\email{boxiong@whut.edu.cn}
\affiliation{Department of Physics, Wuhan University of Technology, Wuhan 430070, China}

\date{\today}

\begin{abstract}
Matrix product operator (MPO) serves as a fundamental component in tensor network simulations of quantum many-body dynamics. We employ an MPO construction that introduces additional propagation channels to embed both periodic boundary conditions and finite-range couplings directly into an open boundary MPO. We apply this construction within the time-dependent variational principle (TDVP) framework to simulate quench dynamics in a spin-1/2 chain with finite-range interactions, and benchmark the results numerically against the fourth-order Runge-Kutta method, finding excellent agreement for both single-body and two-body observables. The approach offers a practical route for tensor network simulations of many-body dynamics in periodic finite-range systems.
\end{abstract}

\maketitle
\section{Introduction}

Long-range (LR) interactions in spin-chain systems play a crucial role in stabilizing exotic quantum phases, such as long-range magnetic order \cite{Koffel2012,Laflorencie2005}, topological phases \cite{Vodola2014}, and discrete time crystals \cite{Kozin2019}. These emergent phases have stimulated intense interest in quantum many-body physics governed by LR couplings, encompassing ground-state phase diagrams \cite{Saadatmand2018,Block2022}, dynamical quantum phase transitions \cite{Halimeh2017,Bojan2018}, and far-from-equilibrium dynamics \cite{Zeiher2017}. Experimentally, rapid progress has established a variety of platforms -- including Rydberg atoms \cite{keesling2019,Scholl2021}, dipolar quantum gases \cite{Bigagli2024,Zhu2025}, polar molecules \cite{li2023,Zhu2025}, quantum gases coupled to optical cavities \cite{Wu2023,Bonifacio2024}, and trapped ions \cite{Britton2012,Monroe2021} -- that realize tunable LR interactions. Exploring the nonequilibrium dynamics in these systems remains a formidable challenge, as it demands theoretical frameworks that faithfully capture the interplay between entanglement growth and long-range correlations.

Matrix product state (MPS) algorithms provide a powerful framework for simulating one-dimensional (1D) quantum many-body systems by efficiently representing finitely entangled states through a product of rank-3 tensors \cite{Vidal2004,Jose2006,Schollwock2011,Paeckel2019,Silvi2019}.
These methods have achieved remarkable success in studies of ground states \cite{White1992,White1993,Schollwock2005} and thermal equilibrium states \cite{Verstraete2004,White2009,Gohlke2023} of many-body systems. For nonequilibrium dynamics, several MPS-based approaches -- such as time-evolving block decimation (TEBD), time-dependent density matrix renormalization group (tDMRG), and the time-dependent variational principle (TDVP) -- have been successfully applied to 1D quantum systems, particularly those governed by nearest-neighbor interactions \cite{Bernier2018,Cai2013,Cui2015,Jaschke2019}.
However, because both TEBD and tDMRG rely on a Suzuki-Trotter decomposition of the time-evolution operator, the bond dimension of the required matrix product operator (MPO) grows exponentially with the interaction range, rendering these methods practically infeasible for systems with long-range couplings. 
Recently, an alternative TEBD-based technique for constructing MPOs of exponentials of nonlocal spin operators has made significant progress in simulating LR interactions \cite{Catalano2025}. Nevertheless, the maximum bond dimension of the resulting MPO still scales as $2^r$, where $r$ is the interaction range, so that for sufficiently large $r$ the approach becomes impractical for truly long-range systems.
The essential insight of TDVP is to project the Schr{\"o}dinger equation onto the tangent space of the manifold formed by MPS of fixed bond dimension. By employing a Lie-Trotter splitting of the tangent-space projector, the evolution is recast as a sequence of coupled differential equations that propagate individual MPS tensors forward and backward in time \cite{Paeckel2019}. These equations can be solved straightforwardly using Krylov-Arnoldi methods. A decisive advantage of TDVP is that the long-range Hamiltonian can be encoded as a compact MPO using finite automaton constructions \cite{Crosswhite2008,Hubig2017,Paeckel2017}; consequently, TDVP can intrinsically handle systems with algebraically decaying interactions. In contrast, approaches that directly construct an MPO representation of the time-evolution operator (such as the $W^{\text{II}}$ method) suffer from a larger time-step error. Because one-site TDVP conserves both the energy and the norm of the state exactly, it yields a more stable long-time evolution \cite{Paeckel2019}.

Despite these strengths, simulating realistic systems that combine finitely long-range interactions with complex geometries remains challenging -- a prominent example being the inclusion of periodic boundary conditions (PBC).
Under PBC, the Hamiltonian MPO must encode long-range couplings that wrap around the ring, which can significantly increase the bond dimension and complicate the MPO structure.
Standard strategies for constructing a periodic MPO often rely on a finite-automaton representation followed by a global trace operation that enforces periodicity \cite{Pirvu2010,Solanki2023}. While such trace-based constructions are widely used for computing ground states or thermal equilibrium properties, they are poorly suited for time evolution when gauge/translational invariance is broken.
The global trace typically in real time evolution destroys the canonical gauge of the local tensors, which compromises the accuracy of the Lie-Trotter splitting and can lead to spurious non-unitary effects or a loss of norm/energy conservation over long times.
Moreover, when long-range interactions are present, the trace-based algorithm often inflates the bond dimension of the translational invariant MPO, significantly slowing down the Krylov-Arnoldi solver at each time step. These issues make it highly desirable to develop efficient factorization schemes for Hamiltonian MPOs that are both compact and fully compatible with the tangent-space projection of TDVP. Such schemes are essential to faithfully probe the nonequilibrium dynamics of long-range systems under PBC, and they constitute the central methodological motivation of this work.
In this work, we develop a systematic MPO construction that introduces additional propagation channels for studying the nonequilibrium dynamics of finite-range interacting spin systems with periodic boundary conditions. This construction can be systematically extended as the interaction range increases. We combine this explicit MPO representation with TDVP to simulate quench dynamics in a spin-1/2 chain with finite-range interactions and benchmark the results numerically against the fourth-order Runge-Kutta (RK4) method, finding excellent agreement for both single-body and two-body observables. Further, we examine how the computational cost depends on the interaction range and system size. This approach offers a practical route for tensor network simulations of many-body dynamics in finite-range systems on periodic lattices.

The remainder of this paper is organized as follows. Section \ref{se2} presents the construction of MPOs for open and periodic boundary conditions, including their extension to finite-range interactions, and introduces the TDVP scheme used for the time evolution. Section \ref{se3} benchmarks the accuracy of the proposed approach against RK4 evolution and examines its computational cost as a function of the interaction range and system size. Section \ref{se4} summarizes our main results.





\section{Matrix Product Operator Representation}\label{se2}

\subsection{Open Boundary MPO Construction}

To represent the Hamiltonian with MPS framework, MPO formalism is employed.
For a one-dimensional lattice consisting of $N$ sites, an operator $H$ can be expressed as
\begin{equation}
\begin{aligned}
H & = \sum_{\sigma, \sigma^{\prime}} W^{[1]}W^{[2]} \cdots W^{[N]} \vert \sigma_{1} \cdots \sigma_{N} \rangle \langle \sigma_{1}^{\prime} \cdots \sigma_{N}^{\prime} \vert \\
& = \raisebox{-5.0mm}{\includegraphics[width=0.8\linewidth]{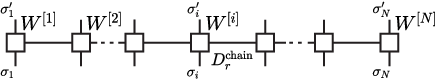}},
\end{aligned}
\end{equation}
where $W^{[i]}$ denotes the local MPO tensor at site $i$. 
Graphically, each local tensor possesses two physical indices $(\sigma_{i}, \sigma^{\prime}_{i})$, which connect to the basis kets $\vert \sigma_{i} \rangle$ and bras $\langle \sigma_{i}^{\prime} \vert$. Under open boundary conditions (OBC), the boundary tensors exhibit distinct structures to ensure a closed system. The initial tensor $W^{[1]}$ is configured as a row vector, whereas the final tensor $W^{[N]}$ is a column vector, fixing their outermost bond dimensions to 1. The intermediate sites share a uniform bulk tensor $W^{[i]} (2\le i \le N-1)$. Adjacent tensors are linked via internal bonds, whose dimension is denoted as $D_{r}^{\rm chain}$.
The construction of these tensors is best illustrated through a specific spin model. The Hamiltonian for the system under consideration is given by
\begin{equation}\label{eq1}
H = \frac{\hbar\Omega}{2}\sum_{i}\sigma_{i}^{x}-\hbar\delta\sum_{i}n_{i}+\sum_{r=1}^{R} V_{r} \sum_{i=1}^{N} n_{i}n_{i+r},
\end{equation}
where $\sigma_{i}^{x}$ is the $x$-Pauli matrix and $n_{i} = (1 + \sigma_{i}^{z})/2$ is the occupation of the excited state in the $i$-th site. $\Omega$ and $\delta$ are treated as transverse and longitudinal fields, respectively. $V_{r}$ determines the interaction strength between the $i$-th and $(i+r)$-th sites.

The simplest case involves exclusively nearest-neighbor interactions ($r=1$). For this configuration, the MPO requires a bond dimension of $D_{r=1}^{\rm chain} = 3$. Specifically, 
\begin{equation} \notag
W^{[1]} = \begin{pmatrix}
    B & V_{1}n & \text{II}
  \end{pmatrix},
\end{equation}	
\begin{equation} \notag
  W^{[2:N-1]} = 
  \begin{pmatrix}
    \text{II} & 0 & 0\\
    n & 0 & 0\\
    B & V_{1}n & \text{II}
  \end{pmatrix},
\end{equation}
\begin{equation} \notag
W^{[N]} = \begin{pmatrix}
    \text{II} & n & B
  \end{pmatrix}^{T},
\end{equation}	
where \text{II} denotes identity matrix and single-site contributions $B = \frac{\Omega}{2}\sigma^{x}-\delta n$.
The MPO representation is specified by the boundary tensors, $W^{[1]}$ and $W^{[N]}$, together with a uniform bulk tensor $W^{[2: N-1]}$. The boundary tensors initiate and terminate the propagation of operators, while the bulk tensor propagates local operators through the virtual bond space and generates the interaction terms contained in Eq.\,(\ref{eq1}).

The mechanism of this representation relies on the sequential multiplication of the local tensors. The boundary tensors, $W^{[1]}$ and $W^{[N]}$, dictate the initiation and termination of the operator expansion. Concurrently, the internal indices of the bulk tensors systematically connect adjacent sites to generate the full many-body interaction terms defined in Eq. (\ref{eq1}). This internal matrix structure offers a direct physical interpretation. The specific rows and columns correspond to distinct operator propagation processes, defined here as propagation channels. Consequently, the bond dimension $D_{r}^{\rm chain}$ mathematically defines the matrix size while physically quantifying the number of these available channels.

Within this nearest-neighbor setup, the onsite operator $B$ directly incorporates local energy contributions without requiring extended propagation. Conversely, the interaction terms necessitate a specific propagation mechanism. The process initiates with the operator $V_{r}n$ at a given site, transfers through the internal matrix indices across the intermediate lattice sites, and terminates upon encountering the operator $n$ at a subsequent site. This configuration is sufficient for nearest-neighbor Interaction. However, longer-range interactions require the correlated operators to span multiple lattice sites before termination. Capturing these extended interactions inherently demands the introduction of additional propagation channels.

\subsection{Finite-Range Interactions MPO}

The nearest-neighbor MPO introduced in the previous section can be systematically generalized to finite-range interactions by enlarging the MPO bond dimension. The central idea is to extend the propagation path of the operator string before it reaches its target site. Compared with the nearest-neighbor case, longer-range interactions require the operator string to traverse additional intermediate lattice sites. During this extended propagation, the operator string merely passes through these sites without physically interacting with them. This is achieved by introducing additional propagation channels into the MPO bond dimension, while preserving the existing interaction channels from the nearest-neighbor construction.

\begin{figure}[htpb]
	\centering
	\includegraphics[scale=1.12]{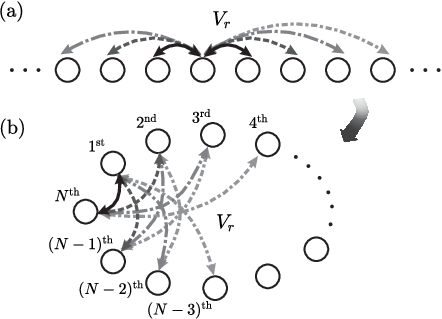}
	\caption{Schematic illustration of finite-range interactions and periodic boundary conditions used for the MPO construction. (a) The 1D chain with OBC, where interactions of different ranges are represented by different line styles. (b) The corresponding periodic chain forms a ring, highlighting the additional interaction terms introduced by the PBC. These interaction patterns determine the additional propagation channels required in the MPO representation.}
    \label{fig1}
\end{figure}

Figure \ref{fig1}\,(a) schematically illustrates the interaction patterns considered for the open boundary chain. Compared with the nearest-neighbor interaction, an interaction of range $r$ spans $r-1$ intermediate lattice sites. Since no interaction is generated on these intermediate sites, they only serve to propagate the interaction toward the target lattice site and are therefore represented by identity operators in the MPO. Consequently, every newly introduced interaction range requires one additional propagation channel, leading to a linear increase of the MPO bond dimension with the maximum interaction distance.

To illustrate the MPO construction, we first consider the second nearest-neighbor interaction $V_{2}n_{i}n_{i+2}$. Compared with the nearest-neighbor case, the interaction initiated by the occupation operator $n_{i}$ must pass through one intermediate lattice site before terminating at site $n_{i+2}$. Accordingly, an additional propagation channel is introduced into the bulk MPO tensor, where the intermediate site is represented solely by the identity operator. In the present construction, this propagation is realized by assigning the identity operator to the $(3,2)$ element of the bulk MPO tensor.
The location of this matrix element is uniquely determined by the propagation rule rather than by an arbitrary assignment. The nearest-neighbor interaction has already established the propagation channel through the $(2,1)$ element. To extend the interaction distance without modifying the existing nearest-neighbor representation, the newly introduced propagation must continue from the second propagation channel instead of creating an independent branch. Consequently, the identity operator is placed at the $(3,2)$ element, establishing a continuous propagation chain. Assigning the identity operator to any other matrix position would either interrupt the existing propagation sequence or alter the nearest-neighbor interaction channels, preventing a systematic extension to longer interaction distances.

The same construction rule can be directly generalized to arbitrary finite-range interactions. For example, incorporating the third nearest-neighbor interaction requires the interaction to propagate across two intermediate lattice sites before reaching its target site. Following exactly the same propagation rule, the propagation chain is extended by one additional step, corresponding to the identity transition at the $(4,3)$ element of the bulk MPO tensor. Repeating this procedure recursively establishes a hierarchical propagation chain, where every newly introduced interaction range extends the existing propagation path by exactly one additional step. Therefore, an interaction of range $r$ requires $r-1$ successive identity propagations, resulting in the linear dependence of the MPO bond dimension on the interaction distance, $D_{r}^{\rm chain} = 3 + (r-1)$, where the first term corresponds to the nearest-neighbor MPO, while the second term represents the additional propagation channels required for finite-range interactions. The resulting bond dimensions for several representative interaction ranges are summarized in Table \ref{table1}.

Following the above construction rule, the bulk MPO tensor describing finite-range interactions can be written in the compact form
\begin{equation} \notag
  W^{[2:N-1]} = 
  \begin{pmatrix}
    \text{II} & 0 & 0 & 0 & \cdots & 0\\
    n & 0 & 0 & 0 & \cdots & 0\\
	0 & \text{II} & 0 & 0 & \cdots & 0 \\
	0 & 0 & \text{II} & 0 & \cdots & 0 \\
	\vdots & \vdots & \vdots & \vdots & \ddots & \vdots \\
    B & V_{1}n & V_{2}n & V_{3}n & \cdots & \text{II}
  \end{pmatrix},
\end{equation}
where the first-site and last-site tensors are obtained from the last row and the first column of the bulk tensor, respectively.

\begin{table}
	\centering
    \setlength{\tabcolsep}{4.5mm}
	\caption{Growth of the MPO bond dimension with interaction range under open and periodic boundary conditions.}
	\label{table1}
	\begin{tabular}{ccc}  
		\toprule
		\toprule
		Boundary & Interaction terms & Dimension of MPO \\
		\midrule
		\multirow{4}{*}{Open} & $V_{1}$ & $3$ \\
		 & $V_{1,2}$ & $4$ \\
		 & $V_{1, 2, 3}$ & $5$\\
		 & $V_{1, \cdots, r}$ & $r + 2$\\
		\multirow{4}{*}{Periodic} & $V_{1}$ & $4 $ \\
		 & $V_{1, 2}$  & $7$ \\
		 & $V_{1, 2, 3}$ & $11$\\
		 & $V_{1, \cdots, r}$ & $ \frac{r^2 + 3r + 4}{2}$\\
		\bottomrule
		\bottomrule
	\end{tabular}
\end{table}

\subsection{Periodic Boundary MPO Construction}

The finite-range MPO constructed above describes interactions under OBC, where all interaction terms are contained within the chain. Under PBC, however, the chain is effectively closed into a ring, and additional interactions connecting the two ends of the lattice must also be incorporated. As illustrated in Fig.\,\ref{fig1}\,(b), for an interaction range $r$, PBC introduce $r$ additional interaction terms to represent periodic interactions. Since these interactions are absent in the open boundary Hamiltonian, they cannot be represented by the existing MPO channels and therefore require additional propagation channels. Meanwhile, these newly introduced channels should remain independent of the open boundary construction so that the original MPO representation is preserved. 

We first consider the nearest-neighbor interaction. Compared with the open boundary case, PBC introduce only one additional interaction between the first and the last lattice sites. To represent this interaction, an additional propagation channel is appended to the first-site MPO tensor by placing the interaction operator $V_{1}n$
at the $(1,x_{1})$ matrix element. Since this interaction exists only between the first and the last lattice sites, it is unnecessary to generate interaction terms on the intermediate sites. Therefore, the corresponding channel in the bulk tensors simply propagates through the identity operator located at the $(x_{1},x_{1})$ element, while the interaction is terminated at the last-site tensor by placing the occupation operator n at the $(x_{1},1)$ element. In this way, the periodic interaction is constructed independently without modifying any channels associated with the open boundary MPO.

The construction can be generalized straightforwardly to longer interactions. For the second nearest-neighbor interaction, two additional periodic interaction terms, namely the first and $(N-1)$-th sites, and the second and $N$-th sites, must be included. Since these interactions originate from different lattice sites and terminate at different target sites, two independent propagation channels are introduced. For the interaction between the first and the $(N-1)$-th lattice sites, the interaction operator $V_{2}n$ is placed at the $(1,x_{2})$ of the first-site tensor, while the corresponding occupation operator is assigned to the $(x_{2},x_{2})$ element of the $(N-1)$-th tensor. The remaining tensors propagate this interaction solely through the identity operators located at the $(x_{2},x_{2})$ elements (the $(x_{2}, 1)$ element of the last site). Similarly, for the interaction between the second and the last lattice sites, the interaction is initiated at the second-site tensor and terminated at the last-site tensor through another independent propagation channel. The remaining tensors again contain only identity operators along this channel, ensuring that the two periodic interaction terms are represented independently and do not interfere with either the open boundary channels or each other.

Following the same construction principle, every newly introduced periodic interaction is assigned an independent propagation channel. Consequently, an interaction range $r$ requires $r$ additional channels to represent the periodic interactions. Combined with the finite-range construction developed in the previous subsection, the bond dimension of the periodic MPO is therefore given by $D_{r}^{\rm ring} = 3 + (r-1) + \frac{r(r+1)}{2}$, which summarized in Table \ref{table1}. It should be noted that, unlike the open boundary construction, the local MPO tensors are no longer identical because the periodic interaction channels originate and terminate at different lattice sites. Consequently, the resulting MPO is nonuniform, while its bond dimension remains independent of the system size and depends only on the maximum interaction range.

According to the construction rules described above, the MPO tensors for PBC can be obtained straightforwardly. For example, the MPO tensors including the third nearest-neighbor interactions are explicitly given by
\vspace{12pt}
\begin{equation} \notag
W^{[1]} = \begin{pNiceMatrix}
	\CodeBefore
	\tikz \draw [fill=gray!60] (1-|5) |- (2-|6) |- cycle ;
	\tikz \draw [fill=gray!35,dashed] (1-|6) |- (3-|8) |- cycle ;
	\tikz \draw [fill=gray!10,dash pattern=on 3pt off 1pt on 0.5pt off 1pt] (1-|8) |- (4-|11) |- cycle ;
	\Body
    B & V_{1}n & V_{2}n & V_{3}n & V_{1}n & V_{2}n & \text{II} & V_{3}n & \text{II} & \text{II} & \text{II}
	\CodeAfter
	\UnderBrace[shorten,yshift=2pt]{1-2}{1-4}{\text{OBC terms}}
	\OverBrace[shorten,yshift=4pt]{1-5}{1-10}{\text{PBC terms}}
  \end{pNiceMatrix},
\end{equation}	
\vspace{2pt}
\begin{equation} \notag
  W^{[2]} = 
  \begin{pNiceMatrix}
	\CodeBefore
	\tikz \draw [fill=gray!60] (5-|5) |- (6-|6) |- cycle ;
	\tikz \draw [fill=gray!35,dashed] (6-|6) |- (7-|7) |- cycle ;
	\tikz \draw [fill=gray!35,dashed] (7-|7) |- (8-|8) |- cycle ;
	\tikz \draw [fill=gray!10,dash pattern=on 3pt off 1pt on 0.5pt off 1pt] (8-|8) |- (9-|9) |- cycle ;
	\tikz \draw [fill=gray!10,dash pattern=on 3pt off 1pt on 0.5pt off 1pt] (9-|9) |- (10-|10) |- cycle ;
	\tikz \draw [fill=gray!10,dash pattern=on 3pt off 1pt on 0.5pt off 1pt] (10-|10) |- (11-|11) |- cycle ;
	\Body
    \text{II} & 0 & 0 & 0 & 0 & 0 & 0 & 0 & 0 & 0 & 0\\
    n & 0 & 0 & 0 & 0 & 0 & 0 & 0 & 0 & 0 & 0\\
	0 & \text{II} & 0 & 0 & 0 & 0 & 0 & 0 & 0 & 0 & 0 \\
	0 & 0 & \text{II} & 0 & 0 & 0 & 0 & 0 & 0 & 0 & 0 \\
	0 & 0 & 0 & 0 & \text{II} & 0 & 0 & 0 & 0 & 0 & 0 \\
	0 & 0 & 0 & 0 & 0 & \text{II} & 0 & 0 & 0 & 0 & 0 \\
	0 & 0 & 0 & 0 & 0 & 0 & V_{2}n & 0 & 0 & 0 & 0 \\
	0 & 0 & 0 & 0 & 0 & 0 & 0 & \text{II} & 0 & 0 & 0 \\
	0 & 0 & 0 & 0 & 0 & 0 & 0 & 0 & V_{3}n & 0 & 0 \\
	0 & 0 & 0 & 0 & 0 & 0 & 0 & 0 & 0 & \text{II} & 0 \\
    B & V_{1}n & V_{2}n & V_{3}n & 0 & 0 & 0 & 0 & 0 & 0 & \text{II}
  \end{pNiceMatrix},
\end{equation}
\begin{equation} \notag
  W^{[3]} = 
  \begin{pNiceMatrix}
	\CodeBefore
	\tikz \draw [fill=gray!60] (5-|5) |- (6-|6) |- cycle ;
	\tikz \draw [fill=gray!35,dashed] (6-|6) |- (7-|7) |- cycle ;
	\tikz \draw [fill=gray!35,dashed] (7-|7) |- (8-|8) |- cycle ;
	\tikz \draw [fill=gray!10,dash pattern=on 3pt off 1pt on 0.5pt off 1pt] (8-|8) |- (9-|9) |- cycle ;
	\tikz \draw [fill=gray!10,dash pattern=on 3pt off 1pt on 0.5pt off 1pt] (9-|9) |- (10-|10) |- cycle ;
	\tikz \draw [fill=gray!10,dash pattern=on 3pt off 1pt on 0.5pt off 1pt] (10-|10) |- (11-|11) |- cycle ;
	\Body
    \text{II} & 0 & 0 & 0 & 0 & 0 & 0 & 0 & 0 & 0 & 0\\
    n & 0 & 0 & 0 & 0 & 0 & 0 & 0 & 0 & 0 & 0\\
	0 & \text{II} & 0 & 0 & 0 & 0 & 0 & 0 & 0 & 0 & 0 \\
	0 & 0 & \text{II} & 0 & 0 & 0 & 0 & 0 & 0 & 0 & 0 \\
	0 & 0 & 0 & 0 & \text{II} & 0 & 0 & 0 & 0 & 0 & 0 \\
	0 & 0 & 0 & 0 & 0 & \text{II} & 0 & 0 & 0 & 0 & 0 \\
	0 & 0 & 0 & 0 & 0 & 0 & \text{II} & 0 & 0 & 0 & 0 \\
	0 & 0 & 0 & 0 & 0 & 0 & 0 & \text{II} & 0 & 0 & 0 \\
	0 & 0 & 0 & 0 & 0 & 0 & 0 & 0 & \text{II} & 0 & 0 \\
	0 & 0 & 0 & 0 & 0 & 0 & 0 & 0 & 0 & V_{3}n & 0 \\
    B & V_{1}n & V_{2}n & V_{3}n & 0 & 0 & 0 & 0 & 0 & 0 & \text{II}
  \end{pNiceMatrix},
\end{equation}
\begin{equation} \notag
  W^{[4:N-3]} = 
  \begin{pNiceMatrix}
	\CodeBefore
	\tikz \draw [fill=gray!60] (5-|5) |- (6-|6) |- cycle ;
	\tikz \draw [fill=gray!35,dashed] (6-|6) |- (7-|7) |- cycle ;
	\tikz \draw [fill=gray!35,dashed] (7-|7) |- (8-|8) |- cycle ;
	\tikz \draw [fill=gray!10,dash pattern=on 3pt off 1pt on 0.5pt off 1pt] (8-|8) |- (9-|9) |- cycle ;
	\tikz \draw [fill=gray!10,dash pattern=on 3pt off 1pt on 0.5pt off 1pt] (9-|9) |- (10-|10) |- cycle ;
	\tikz \draw [fill=gray!10,dash pattern=on 3pt off 1pt on 0.5pt off 1pt] (10-|10) |- (11-|11) |- cycle ;
	\Body
    \text{II} & 0 & 0 & 0 & 0 & 0 & 0 & 0 & 0 & 0 & 0\\
    n & 0 & 0 & 0 & 0 & 0 & 0 & 0 & 0 & 0 & 0\\
	0 & \text{II} & 0 & 0 & 0 & 0 & 0 & 0 & 0 & 0 & 0 \\
	0 & 0 & \text{II} & 0 & 0 & 0 & 0 & 0 & 0 & 0 & 0 \\
	0 & 0 & 0 & 0 & \text{II} & 0 & 0 & 0 & 0 & 0 & 0 \\
	0 & 0 & 0 & 0 & 0 & \text{II} & 0 & 0 & 0 & 0 & 0 \\
	0 & 0 & 0 & 0 & 0 & 0 & \text{II} & 0 & 0 & 0 & 0 \\
	0 & 0 & 0 & 0 & 0 & 0 & 0 & \text{II} & 0 & 0 & 0 \\
	0 & 0 & 0 & 0 & 0 & 0 & 0 & 0 & \text{II} & 0 & 0 \\
	0 & 0 & 0 & 0 & 0 & 0 & 0 & 0 & 0 & \text{II} & 0 \\
    B & V_{1}n & V_{2}n & V_{3}n & 0 & 0 & 0 & 0 & 0 & 0 & \text{II}
  \end{pNiceMatrix},
\end{equation}
\begin{equation} \notag
  W^{[N-2]} = 
  \begin{pNiceMatrix}
	\CodeBefore
	\tikz \draw [fill=gray!60] (5-|5) |- (6-|6) |- cycle ;
	\tikz \draw [fill=gray!35,dashed] (6-|6) |- (7-|7) |- cycle ;
	\tikz \draw [fill=gray!35,dashed] (7-|7) |- (8-|8) |- cycle ;
	\tikz \draw [fill=gray!10,dash pattern=on 3pt off 1pt on 0.5pt off 1pt] (8-|8) |- (9-|9) |- cycle ;
	\tikz \draw [fill=gray!10,dash pattern=on 3pt off 1pt on 0.5pt off 1pt] (9-|9) |- (10-|10) |- cycle ;
	\tikz \draw [fill=gray!10,dash pattern=on 3pt off 1pt on 0.5pt off 1pt] (10-|10) |- (11-|11) |- cycle ;
	\Body
    \text{II} & 0 & 0 & 0 & 0 & 0 & 0 & 0 & 0 & 0 & 0\\
    n & 0 & 0 & 0 & 0 & 0 & 0 & 0 & 0 & 0 & 0\\
	0 & \text{II} & 0 & 0 & 0 & 0 & 0 & 0 & 0 & 0 & 0 \\
	0 & 0 & \text{II} & 0 & 0 & 0 & 0 & 0 & 0 & 0 & 0 \\
	0 & 0 & 0 & 0 & \text{II} & 0 & 0 & 0 & 0 & 0 & 0 \\
	0 & 0 & 0 & 0 & 0 & \text{II} & 0 & 0 & 0 & 0 & 0 \\
	0 & 0 & 0 & 0 & 0 & 0 & \text{II} & 0 & 0 & 0 & 0 \\
	0 & 0 & 0 & 0 & 0 & 0 & 0 & n & 0 & 0 & 0 \\
	0 & 0 & 0 & 0 & 0 & 0 & 0 & 0 & \text{II} & 0 & 0 \\
	0 & 0 & 0 & 0 & 0 & 0 & 0 & 0 & 0 & \text{II} & 0 \\
    B & V_{1}n & V_{2}n & V_{3}n & 0 & 0 & 0 & 0 & 0 & 0 & \text{II}
  \end{pNiceMatrix},
\end{equation}
\begin{equation} \notag
  W^{[N-1]} = 
  \begin{pNiceMatrix}
	\CodeBefore
	\tikz \draw [fill=gray!60] (5-|5) |- (6-|6) |- cycle ;
	\tikz \draw [fill=gray!35,dashed] (6-|6) |- (7-|7) |- cycle ;
	\tikz \draw [fill=gray!35,dashed] (7-|7) |- (8-|8) |- cycle ;
	\tikz \draw [fill=gray!10,dash pattern=on 3pt off 1pt on 0.5pt off 1pt] (8-|8) |- (9-|9) |- cycle ;
	\tikz \draw [fill=gray!10,dash pattern=on 3pt off 1pt on 0.5pt off 1pt] (9-|9) |- (10-|10) |- cycle ;
	\tikz \draw [fill=gray!10,dash pattern=on 3pt off 1pt on 0.5pt off 1pt] (10-|10) |- (11-|11) |- cycle ;
	\Body
    \text{II} & 0 & 0 & 0 & 0 & 0 & 0 & 0 & 0 & 0 & 0\\
    n & 0 & 0 & 0 & 0 & 0 & 0 & 0 & 0 & 0 & 0\\
	0 & \text{II} & 0 & 0 & 0 & 0 & 0 & 0 & 0 & 0 & 0 \\
	0 & 0 & \text{II} & 0 & 0 & 0 & 0 & 0 & 0 & 0 & 0 \\
	0 & 0 & 0 & 0 & \text{II} & 0 & 0 & 0 & 0 & 0 & 0 \\
	0 & 0 & 0 & 0 & 0 & n & 0 & 0 & 0 & 0 & 0 \\
	0 & 0 & 0 & 0 & 0 & 0 & \text{II} & 0 & 0 & 0 & 0 \\
	0 & 0 & 0 & 0 & 0 & 0 & 0 & \text{II} & 0 & 0 & 0 \\
	0 & 0 & 0 & 0 & 0 & 0 & 0 & 0 & n & 0 & 0 \\
	0 & 0 & 0 & 0 & 0 & 0 & 0 & 0 & 0 & \text{II} & 0 \\
    B & V_{1}n & V_{2}n & V_{3}n & 0 & 0 & 0 & 0 & 0 & 0 & \text{II}
  \end{pNiceMatrix},
\end{equation}
\begin{equation} \notag
W^{[N]} = \begin{pNiceMatrix}
	\CodeBefore
	\tikz \draw [fill=gray!60] (1-|5) |- (2-|6) |- cycle ;
	\tikz \draw [fill=gray!35,dashed] (1-|6) |- (3-|8) |- cycle ;
	\tikz \draw [fill=gray!10,dash pattern=on 3pt off 1pt on 0.5pt off 1pt] (1-|8) |- (4-|11) |- cycle ;
	\Body
    \text{II} & n & 0 & 0 & n & \text{II} & n & \text{II} & \text{II} & n & B
  \end{pNiceMatrix}^{T},
\end{equation}	
where the additional matrix elements associated with the periodic channels describe only the periodic interactions, while the original matrix structure of the open boundary MPO remains unchanged. 
For clarity, the matrix elements introduced by the periodic first, second, and third nearest-neighbor interactions are highlighted using solid, dashed, and dash-dotted gray boxes, respectively, consistent with the graphical representation in Fig.\ref{fig1}\,(b). These annotations explicitly distinguish the additional periodic interaction channels from the original open boundary MPO and illustrate how periodic interactions of different ranges are incorporated into the local tensors.

The construction for longer interaction ranges follows the same principle by introducing the corresponding propagation channels associated with the additional periodic interactions.
Unlike the OBC, the local MPO tensors are no longer identical because the periodic interactions do not affect all sites uniformly.
Nevertheless, this nonuniformity is confined to the boundary region. For a maximum interaction range $r$, only the tensors associated with the first $r$ sites and the last $r$ lattice sites require individual constructions, whereas all tensors from $(r+1)$-th to $(N-r)$-th sites share the same bulk MPO tensor.
Therefore, the number of nonuniform tensors in the periodic MPO depends on the maximum interaction range, while the remaining tensors retain an identical structure. 

\subsection{Time-dependent Variational Principle}

The simulation of nonequilibrium quantum dynamics has become one of the central applications of tensor network methods \cite{Schollwock2011,Paeckel2019}. Among the available real-time evolution algorithms, TEBD and tDMRG have been successfully applied to a wide variety of one-dimensional quantum systems \cite{Vidal2003,Vidal2004,Daley2004,White2004}. These approaches typically rely on a Suzuki-Trotter decomposition of the time-evolution operator into a sequence of local gates, making them particularly efficient for short-range Hamiltonians. However, for finite-range or long-range interactions, the decomposition requires a rapidly increasing number of noncommuting terms, resulting in a considerable increase in computational complexity and Trotter errors. In contrast, TDVP performs the time evolution directly within the variational manifold of MPS without requiring any Trotter decomposition \cite{Haegeman2011,Haegeman2016}. Since the Hamiltonian enters the evolution only through its MPO representation, TDVP naturally accommodates arbitrary finite-range interactions once an appropriate MPO is available. This feature makes it particularly suitable for the periodic finite-range MPO developed in the present work.

The real-time evolution of a closed quantum system is governed by the time-dependent Schr{\"o}dinger equation $(\hbar = 1)$
\begin{equation}
	i\frac{\partial}{\partial t}|\psi\rangle = H \vert \psi\rangle,
\end{equation}
where the Hamiltonian $H$ is represented by the periodic finite-range MPO introduced in the previous subsections. Within the tensor network framework, the many-body wave function is expressed in the MPS form, 
\begin{equation}
	\vert \psi\rangle = \sum_{\sigma} A^{[1]} A^{[2]} \cdots A^{[N]} \vert \sigma_{1}\sigma_{2}\cdots\sigma_{N}\rangle,
\end{equation}
where $A^{[i]}$ denotes the local tensor at the $i$-th site with physical index $\sigma_{i}$ and bond dimensions $D_{i-1}$ and $D_{i}$. The set of basis states $\vert \sigma_{1}\sigma_{2}\cdots\sigma_{N}\rangle$ spans the complete Hilbert space of the spin chain. Instead of evolving the wave function in the full Hilbert space, TDVP projects the Schr{\"o}dinger equation onto the tangent space of the MPS manifold,
\begin{equation}
	\frac{\partial}{\partial t} \vert \psi\rangle = -i \mathcal{P}_{T} H \vert \psi\rangle,
\end{equation}
where $\mathcal{P}_T$ denotes the projector onto the tangent space. Specifically,
\begin{equation}
    \mathcal{P}_{T} = \sum_{i=1}^{N} P_{L}^{[1:i-1]} \otimes \text{II}_{i} \otimes P_{R}^{[i+1:N]} - \sum_{i=1}^{N-1} P_{L}^{[1:i]} \otimes P_{R}^{[i+1:N]},
\end{equation}
with $P_{L}$ and $P_{R}$ denoting the projectors onto the gauge-fixed left- and right-normalized MPS subspaces, respectively.
This projection constrains the evolution to remain within the variational MPS manifold.
Since the Hamiltonian appears only through the effective MPO acting on the local tensors, the finite-range periodic MPO constructed in previous subsections can be incorporated directly into the TDVP evolution without modifying the variational algorithm itself. In this way, all finite-range interactions and periodic boundary effects are treated consistently through the proposed MPO representation.

In this work, the initial state is chosen as the ground state with all sites in $\vert 0 \rangle$, where the local basis is $\{\vert 0 \rangle$, $\vert 1 \rangle\}$. In the MPS representation, this state is constructed by taking the local tensor $A^{[i]}$ with components $1$ for $\vert 0 \rangle$ and $0$ for $\vert 1 \rangle$, and with bond dimensions $D_{i-1} = D_{i} = 1$.
During the quench dynamics, quantum correlations develop continuously, leading to a gradual increase of the entanglement.
To capture this entanglement growth efficiently, we employ the two-site TDVP algorithm early in the evolution, allowing the MPS bond dimension to increase dynamically until a prescribed maximum value $D_{\rm max}$ is reached. After the bond dimension saturates at the chosen upper limit, the evolution is switched to the more efficient one-site TDVP algorithm. 
This combined two-site and one-site TDVP strategy thus enables adaptive bond dimension growth during the initial entanglement buildup and efficient long-time evolution after the desired variational space has been established.
All numerical simulations presented in this work are performed using a self-developed Fortran code.

\section{Numerical Validation and Analysis}\label{se3}


To evaluate the numerical accuracy and computational efficiency of the proposed framework, the previously established finite-range MPO representations are combined with the TDVP algorithm to investigate the nonequilibrium dynamics. The system is driven out of equilibrium through a typical three-stage quench protocol. To provide a precise description of this evolution, the time-dependent Rabi frequency $\Omega(t)$ and detuning $\delta(t)$ are explicitly formulated using piecewise continuous functions.
The dynamic sweep of the parameters is defined as follows:
\begin{align}
	\Omega(t) & =
	\begin{cases}
	k_{1}t, & 0 \le t < t_1\\
	\Omega_{\rm max}, & t_1 \le t < t_2\\
	-k_{2}t, & t_2 \le t \le t_3,\end{cases}
\\
	\delta(t) & =
	\begin{cases}
	\delta_{i}, & 0 \le t < t_1\\
	\delta_{i} + k_{3}t, & t_1 \le t < t_2\\
	\delta_{f}, & t_2 \le t \le t_3.
    \end{cases}
\end{align}
where $k_{1} = \Omega_{\rm max}/t_{\rm rise}$, $k_{2} = \Omega_{\rm max}/t_{\rm fall}$, and $k_{3} = (\delta_{f} - \delta_{i})/t_{\rm sweep}$. The $\Omega_{\max}$ denotes the maximum Rabi frequency. The $\delta_i$ and $\delta_f$ represent the initial and the final value of the detuning, respectively. In this work, $t_{\rm rise} = t_{\rm fall}= 0.1 \mu s$ and $t_{\rm sweep} = 0.5 \mu s$.
Such parameter sweeps are widely implemented in researches, particularly within programmable Rydberg atomic quantum simulators \cite{Bernien2017,Lienhard2018,Guardado2018,Ebadi2021}. 

The nonequilibrium dynamics is characterized by the average occupation number and the connected correlation functions. The average occupation number is defined as
\begin{equation}
n(t)=\frac{1}{N}\sum_{i=1}^{N}\langle n_i(t)\rangle,
\end{equation}
which describes the average population of the excited state during the dynamical evolution. 
The connected correlation function between two sites separated by a distance $r$ is given by
\begin{equation}
C_{r}(t)=\frac{1}{N_{r}}\sum_{i=1}^{N_{r}}\left[\langle n_i(t)n_{i+r}(t)\rangle-\langle n_i(t)\rangle\langle n_{i+r}(t)\rangle\right],
\end{equation}
where $r$ denotes the distance between two sites. 
In this work, the results are obtained for a 1D chain including the nearest-, next-nearest-, and third-nearest-neighbor interactions, with interaction strengths $V_1 = 2\pi \times 3.2 \, \rm MHz$, $V_2 = 2\pi \times 0.4 \, \rm MHz$, and $V_3 = 2\pi \times 0.12 \, \rm MHz$.

\subsection{Benchmark of the TDVP dynamics}

To verify the reliability of the finite-range MPO representation combined with the TDVP algorithm, we first compare the TDVP results with the conventional RK4 method. The calculations are performed for a periodic chain with $N=16$ sites, including interactions up to the third-nearest-neighbor terms. The same quench protocol described above is employed for both methods, and the results obtained from RK4 are used as the reference solution.

\begin{figure}[htpb]
	\centering
	\includegraphics[scale=0.97]{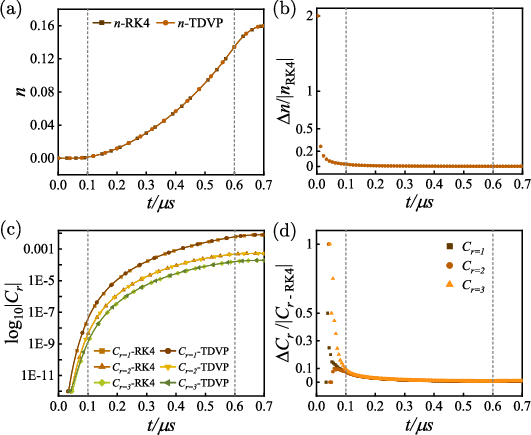}
	\caption{Numerical benchmark of TDVP against RK4 method for quantum dynamical evolution. (a) Time dependence of the average occupation number $n$. (b) Relative error of the occupation number. (c) Time evolution of the absolute values of three spatial correlation functions $C_{r} (r = 1, 2, 3)$ displayed on a base-10 logarithmic vertical axis. (d) Relative errors of the correlation functions versus evolution time. The gray dashed lines indicate the transitions between different stages of the quench protocol.}
    \label{fig2}
\end{figure}

Figure\,\ref{fig2}\,(a) compares the time evolution of the average occupation obtained from TDVP and RK4. The two curves are nearly indistinguishable throughout the entire evolution, indicating that the MPO representation and the TDVP time evolution accurately reproduce the dynamical behavior. To quantify the deviation between the two methods, we define the relative error of the occupation as
\begin{equation}
\frac{\Delta n}{|n_{\rm RK4}|} = \frac{|n_{\rm TDVP}-n_{\rm RK4}|}{|n_{\rm RK4}|} .
\end{equation}

As shown in Fig.\,\ref{fig2}\,(b), the relative error is relatively large only at the very beginning of the evolution, where the occupation remains extremely small ($n\sim10^{-9}$). In this regime, even a small absolute deviation between the two numerical methods can lead to an amplified relative error. 
Once the occupation becomes appreciably populated, the relative error rapidly decreases. During the main evolution stage from $0.1~\mu s$ to $0.6 \mu s$, the relative error continuously decreases with increasing occupation and remains within $1\%$, with an average value of approximately $0.58\%$. In the final stage of the evolution, the deviation slightly increases as the driving protocol changes, but remains below $0.15\%$. These results demonstrate the excellent agreement between TDVP and RK4 for the single-site observable throughout the complete dynamical process.

We further examine the connected correlation functions to evaluate whether the MPO-based TDVP evolution can accurately capture the development of spatial correlations. 
Since the magnitudes of the correlations at different distances differ by several orders of magnitude, the absolute values of the correlation functions are presented on a logarithmic scale in Fig.\,\ref{fig2}\,(c). The TDVP and RK4 results show excellent agreement for the nearest-, second-, and third-nearest-neighbor correlations. Meanwhile, the figure also reveals the characteristic propagation of correlations during the quench process. The nearest-neighbor correlation develops first, whereas correlations at larger distances emerge progressively later due to the finite propagation time required for long-range correlations to establish.

The corresponding relative errors are defined as
\begin{equation}
\frac{\Delta C_{r}}{|C_{r\text{-}\rm RK4}|} = \frac{|C_{r\text{-}\rm TDVP}-C_{r\text{-}\rm RK4}|}{|C_{r\text{-}\rm RK4}|}.
\end{equation}

As shown in Fig.\,\ref{fig2}\,(d), the relative errors exhibit larger fluctuations only during the initial stage when the correlations are still extremely small. After the correlations increase to approximately the order of $10^{-6}$, the relative errors for all interaction distances decrease below $5\%$ and the average relative error is below $1.5\%$ throughout the subsequent evolution. During the evolution, the minimum errors achieved for the nearest-, second-, and third-nearest-neighbor correlations are $0.63\%$, $0.79\%$, and $0.82\%$, respectively.

Overall, the benchmark demonstrates that the finite-range MPO constructed in this work can be reliably incorporated into the TDVP framework. The TDVP evolution reproduces the RK4 results with high accuracy for both local observables and spatial correlations, demonstrating the reliability and the numerical stability of the proposed MPO representation for nonequilibrium dynamics simulations.

\subsection{Computational efficiency and scalability}

Having established the agreement between the MPO-based TDVP calculation and the RK4 benchmark in Sec.\,\ref{se3}\,A, we next examine the computational cost of the TDVP simulations as a function of the interaction range and system size. All calculations use the same dynamical protocol and numerical parameters described above, with a total evolution time of $0.7 \mu s$.

\begin{figure}[htpb]
	\centering
	\includegraphics[scale=0.97]{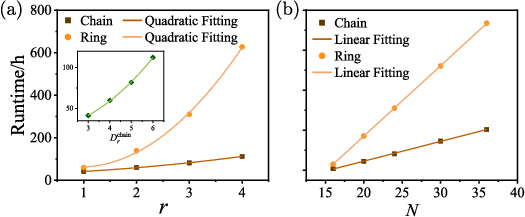}
	\caption{Computational cost of the TDVP simulations as a function of the interaction range and system size. OBC and PBC results are represented by brown filled squares and orange filled circles, respectively. (a) Runtime as a function of the maximum interaction range $r$ for $N=24$, with second-order polynomial fits shown as solid lines. The inset shows the OBC runtime as a function of the corresponding MPO bond dimension $D_{r}^{\rm OBC}$, where green diamonds and a green solid line denote the data and second-order polynomial fit, respectively. (b) Runtime as a function of system size $N$ for $r=3$, with linear fits shown as solid lines.}
    \label{fig3}
\end{figure}

Figure\,\ref{fig3}\,(a) shows the runtime as a function of the maximum interaction range $r$ for a fixed system size $N = 24$. The inset further shows the OBC runtime as a function of the corresponding MPO bond dimension, providing a direct connection between the computational cost and the MPO representation constructed in Sec.\,\ref{se2}.

For OBC, the runtime increases with the interaction range and exhibits a nonlinear dependence over the range considered [see brown filled squares in Fig.\,\ref{fig3}\,(a)]. This behavior is closely related to the growth of the MPO bond dimension, $D_{r}^{\rm{chain}}=r+2$. As $r$ increases, additional channels are introduced to represent the longer-range interactions, thereby increasing the dimension of the MPO used in the TDVP evolution.
The inset of Fig.\,\ref{fig3}\,(a) shows that the runtime also exhibits a nonlinear dependence on $D_{r}^{\rm chain}$, with a second-order polynomial providing a better description of the four data points than a linear fit. Thus, the increase in computational cost with interaction range can be understood as a consequence of the combined effects of the increasing MPO dimension and the resulting contraction cost.

The computational overhead becomes substantially larger for PBC [see  orange filled circles in Fig.\,\ref{fig3}\,(a)]. In this case, the MPO dimension is $D_{r}^{\rm ring} = \frac{r^2+3r+4}{2}$, which grows more rapidly with $r$ than the linear dependence for OBC.
The additional channels required to represent the periodic contributions therefore lead to a more rapid increase in the MPO dimension as longer-range interactions are included.
Consistent with this behavior, the PBC runtime increases markedly with $r$, particularly for $r=3$ and $r=4$. The second-order polynomial fit captures the nonlinear trend over the interaction ranges considered. The stronger increase in PBC runtime can thus be related to the more rapid growth of the MPO representation required by the periodic construction.

We further investigate the dependence on system size in Fig.\,\ref{fig3}\,(b) by fixing the interaction range at $r=3$. The corresponding MPO dimensions are then fixed at $D_{r}^{\rm chain}=5$ and $D_{r}^{\rm ring}=11$, allowing the effect of $N$ to be examined independently of the interaction-range dependence. For both boundary conditions, the runtime increases approximately linearly with system size over the range studied. The larger slope observed for PBC reflects its higher computational cost at the same interaction range, consistent with the larger MPO dimension required by the periodic construction.

Overall, the results establish a clear connection between the MPO representation and the computational cost of the TDVP simulations. Increasing the interaction range introduces additional MPO channels and increases the bond dimension, with a substantially stronger increase for PBC. At fixed interaction range, the runtime grows approximately linearly with system size. The polynomial and linear fits in Fig.\,\ref{fig3} are used to characterize the numerical trends over the parameter range studied.


\section{Conclusion}\label{se4}

In summary, we develop an MPO framework for representing finite-range spin Hamiltonians under both open and periodic boundary conditions and integrate the resulting MPOs into a TDVP-based time-evolution scheme. For OBC, longer-range interactions can be incorporated systematically by introducing additional MPO channels, leading to a bond dimension $D_{r}^{\rm chain}=r+2$. For PBC, the periodic contributions require additional channels associated with different interaction ranges, giving the bond dimension $D_{r}^{\rm ring} = \frac{r^2+3r+4}{2}$.
This construction provides a systematic way to extend MPO representations from short-range to longer-range interactions while retaining a form suitable for tensor-network time evolution.

The accuracy of the resulting TDVP implementation was validated against RK4 evolution for a periodic system with interactions up to the third neighbor. The TDVP results show excellent agreement with RK4 for both single-site occupations and connected correlations, with an average relative error of approximately $0.58\%$ for the occupation and below $1.5\%$ for the correlations during the subsequent evolution. It indicates that the proposed MPO-TDVP approach accurately captures both local observables and longer-range correlations in nonequilibrium dynamics.
We further examined the computational cost as a function of interaction range and system size. The runtime increases with the interaction range, with a more pronounced increase for PBC, consistent with the more rapid growth of the MPO bond dimension in the periodic construction. At fixed interaction range, the runtime increases approximately linearly with system size over the range studied. These results highlight the computational overhead associated with representing longer-range periodic interactions and provide a practical characterization of the cost of the proposed MPO-TDVP approach.

The results establish a consistent connection between the MPO construction, its bond-dimension growth, and the computational cost of TDVP simulations. The framework therefore provides a practical approach for studying nonequilibrium dynamics of finite-range spin systems with periodic boundary conditions, while allowing systematic extension to longer interaction ranges.

\section*{Acknowledgment}
We would like to thank Jun-Hui Zheng for valuable discussions.
This work is supported by the National Natural Science Foundation of China under Grants No.12075175 and No.12575028.

\bibliography{tdvp}

\end{document}